\documentclass{elsarticle}

\usepackage{lineno,hyperref}

\journal{arxiv.org}

\usepackage[margin=2.5cm]{geometry}
\usepackage{amsmath, amsthm, amssymb}
\usepackage{algorithm}
\usepackage{algpseudocode}
\usepackage{tikz}
\usetikzlibrary{decorations.pathreplacing,calc}

\newcommand{\R}{{\rm R}}

\algnewcommand\algorithmicinput{\textbf{Input:}}
\algnewcommand\algorithmicoutput{\textbf{Output:}}
\algnewcommand\Input{\item[\algorithmicinput]}
\algnewcommand\Output{\item[\algorithmicoutput]}

\newcommand{\tikzmark}[1]{\tikz[overlay,remember picture] \node (#1) {};}

\newcommand*{\AddNote}[4]{%
  \begin{tikzpicture}[overlay, remember picture]
    \draw [decoration={brace,amplitude=0.5em},decorate,ultra thick]
        ($(#3)!(#1.north)!($(#3)-(0,1)$)$) --  
        ($(#3)!(#2.south)!($(#3)-(0,1)$)$)
           node [align=center, text width=3.5cm, pos=0.5, anchor=west] {#4};
  \end{tikzpicture}
}

\begin{document}

\begin{frontmatter}

\title{Efficient search of optimal Flower Constellations}

\author{Mart\'\i n Avenda\~no\corref{correspondingauthor}}
\cortext[correspondingauthor]{Corresponding author}
\address{CUD-AGM (Zaragoza), Crtra Huesca s/n, Zaragoza, 50090, Spain}
\ead{avendano@unizar.es}
\author{David Arnas}
\ead{arnas@mit.edu}
\author{Richard Linares}
\ead{linaresr@mit.edu}
\author{Miles Lifson}
\ead{mlifson@mit.edu}
\address{Massachusetts Institute of Technology, Cambridge, MA, 02139, USA}

\begin{abstract}
We derive an analytical closed expression to compute the
minimum distance (quantified by the angle of separation measured from the
center of the Earth) between any two satellites located at the same
altitude and in circular orbits.  We also exploit several properties of
Flower Constellations (FCs) that, combined with our formula for the
distance, give an efficient method to compute the minimum angular distance
between satellites, for all possible FCs with up to a given number of
satellites.
\end{abstract}


\end{frontmatter}

\section{Introduction}

A reasonable slotting system for the Low Earth Orbit (LEO) region can be
obtained from a series of concentric Flower Constellations (FCs) with
circular orbits, but orbit inclination and number of satellites varying
between layers~\cite{stmfc,leemorav16}. The main constraint imposed to
these FCs is
that their dynamics must guarantee that no collisions can occur (i.e.~the
satellites are always separated by a given minimum distance) at any
instant of time.
The distance between satellites that belong to one such FC depend directly
on the altitude of the layer (orbit radius) and the angle of separation
between them. For this reason, and to allow for a layer-independent design,
we decided to evaluate FCs based only on two criteria: number of satellites
$N_{sat}$, and minimum angular separation $\alpha_{min}$ between any two
satellites during a complete orbital period.
In this paper, we propose an efficient way to tabulate all possible FCs
with up to a maximum number of satellites, including columns for each of
the parameters that define the FC and two extra columns for $N_{sat}$ and
$\alpha_{min}$.

\bigskip

A FC, more precisely, a 2D Lattice Flower Constellation~\cite{2dlfc},
is defined by three integer parameters: the number of orbits
$N_o\geq 1$, the number of satellites per orbit $N_{so}\geq 1$, and a
configuration number $0\leq N_c<N_o$. It also requires the orbital elements
of a reference satellite $a,e,i,\omega,\Omega,M_0$, which are six real
numbers. In our case, we are interested in FCs with circular orbits, so the
eccentricity $e$ and the argument of the perigee $\omega$ can both be
assumed to be zero. Since the two evaluation criteria for FCs,
namely $N_{sat}$ and $\alpha_{min}$,
are invariant with respect to rotations about the axis of rotation of the
Earth and shifts of the time-scale, we can also assume that $\Omega$ and
$M_0$ are zero.
Finally, we have agreed on using the separation angle $\alpha_{min}$ instead
of the actual distance between satellites, so we have no dependence on the
orbit radius $a$ either. This means that the only real (continuous)
parameter of interest is the orbit inclination $i\in[0,\pi]$, which we
propose to discretize in small increments, so that we reduce the problem
to studying a finite number of possible inclinations.

\bigskip

We prove that the number of FCs with $N_{sat}\leq k$ is between $k^2/2$ and
$k^2$. In particular, a database with all such FCs with the inclination
discretized in $l$ possible values will have at most $k^2\cdot l$ rows.
Each row contains the integers $N_o,N_{so},N_c,N_{sat}$, which are all
bounded by $k$ and the reals $i,\alpha_{min}$, which are both between $0$
and $\pi$. Each integer occupies at most $\lceil\log_{10}(k)\rceil+1$
characters, and each real occupies $10$ characters assuming one for the
integer part, one for the decimal point, and eight for the fractional part.
If the values are separated by one space character, and each line is
terminated by a new-line character, we have rows of length
$4\lceil\log_{10}(k)\rceil+30$ characters.
This means that the whole database fits into a file of size
\[ k^2\cdot l\cdot (4\lceil\log_{10}(k)\rceil+30). \]
For $k=10^4$ and $l=180$, this gives about $828$ gigabytes, which is
feasible with current technology.

\bigskip

Computing the value of $\alpha_{min}$ for a given FC is not an easy task.
For each pair of satellites in the FC, we have to propagate their position
during an orbital period, find out when they are at their closest distance,
and then take the minimum of all those values. To help with this process,
we have proven a series of results that reduces the computation time
significantly. The first is a simple formula that, given the orbital
parameters of two satellites moving in circular orbits of the same radius,
provides the value of their minimum angular distance. No propagation is
needed. The second is a theorem that shows that only the distances between
the reference satellite of the FC and the others have to be computed. This
reduces the number of pairs from $N_{sat}\cdot (N_{sat}-1)/2$ to
$N_{sat}-1$.
A third result reduces the number of pairs to $\lfloor N_{sat}/2\rfloor$.
Combining these three results, the amount of computation that has to be
done per FC is about $C\cdot N_{sat}/2$, where $C$ is the number of
floating-point operations in our first formula. Summing over all possible
FCs with
$N_{sat}\leq k$ and $l$~possible values for the inclination, we get about
$C/4\cdot k^3\cdot l$ operations.

\bigskip

Our final result is a characterization of all FCs with satellite collisions.
In general, the condition for having collisions depends on the inclination.
However, we found a family of FCs (those with both $N_{o}$ and $N_{so}+N_c$
even integers), accounting for $25\%$ of the search space, that can be
proven to have collisions regardless of the value of the inclination.
Pruning those FCs out from our database, since they clearly have
$\alpha_{min}=0$, reduces the computation time and the size of the database
by a factor of $0.75$.

\bigskip

We have implemented all these techniques in OpenCL, and a preliminary
result (on a laptop with an Intel Gen9 HD Graphics NEO GPU) shows that for
$k=10^4$ and $l=180$, the $0.75\cdot 828 = 621$ gigabytes database can
be computed in $45.5$ hours. On an Nvidia Tesla K40c GPU, this time
reduces to $3.35$ hours, disregarding the I/O time required to save th
data to the hard drive.

\bigskip

The techniques developed in Section~\ref{sec3} can be extrapolated
easily to the more general case of 2D Necklace Flower Constellations~\cite{arnas18}.

\section{Fast computation of the minimum angular distance}\label{sec2}

In this section, we derive a closed formula (no propagation needed) to
compute the minimum angular distance between two satellites $Sat_1$ and
$Sat_2$, moving in circular orbits of the same radius $R$.
The orbital elements of these satellites are
\[
  \begin{aligned}
    Sat_1 & \rightsquigarrow (a=R, e=0, i_1, \omega=0, \Omega_1, M_{01}) \\
    Sat_2 & \rightsquigarrow (a=R, e=0, i_2, \omega=0, \Omega_2, M_{02})
  \end{aligned}
\]
where $i_1,i_2\in[0,\pi]$ and $\Omega_1,\Omega_2,M_{01},M_{02}\in[0,2\pi]$.
The mean motion and period of the satellites are
\[
  T = 2\pi\sqrt{\frac{R^3}{\mu}}\qquad
  n = \frac{2\pi}{T} = \sqrt{\frac{\mu}{R^3}}
\]
where $\mu\approx 3.986\cdot 10^{14}m^3/s^2$ is the standard gravitational
parameter of the Earth.
The positions $\vec{r}_1(t)$ and $\vec{r}_2(t)$ of the satellites are given
by
\[
  \begin{aligned}
    \vec{r}_1(t) & = \R_z(\Omega_1)\R_x(i_1)\R_z(M_{01})
       \left(\begin{array}{c}R\cos(nt)\\ R\sin(nt) \\ 0\end{array}
       \right) \\
    \vec{r}_2(t) & = \R_z(\Omega_2)\R_x(i_2)\R_z(M_{02})
       \left(\begin{array}{c}R\cos(nt)\\ R\sin(nt) \\ 0\end{array}
       \right) \\
  \end{aligned}
\]
in the ECI (Earth Centered Inertial) reference frame. The unit vectors
$\hat{r}_1(t)$ and $\hat{r}_2(t)$ are given by the same expressions,
but setting $R=1$. The convention used here for the rotation matrices is
\[
  \R_x(\alpha)=\left(
    \begin{array}{ccc}
     1 & 0 & 0 \\
     0 & \cos(\alpha) & -\sin(\alpha) \\
     0 & \sin(\alpha) &  \cos(\alpha)
    \end{array}\right)
  \qquad
  \R_z(\alpha)=\left(
    \begin{array}{ccc}
     \cos(\alpha) & -\sin(\alpha) & 0 \\
     \sin(\alpha) &  \cos(\alpha) & 0 \\
     0 & 0 & 1
    \end{array}\right)
\]
for any angle $\alpha\in[0,2\pi]$.
At any instant of time $t$, the angle $\gamma(t)$ between $\hat{r}_1(t)$
and $\hat{r}_2(t)$ satisfies
\[
  \begin{aligned}
    \cos&(\gamma(t)) = \hat{r}_2^T(t)\hat{r}_1(t) = \\
   & = (\cos(nt),\; \sin(nt),\; 0) \R_z(-M_{02})\R_x(-i_2)\R_z(-\Omega_2)
       \R_z(\Omega_1)\R_x(i_1)\R_z(M_{01})\left(\begin{array}{c}
         \cos(nt)\\ \sin(nt) \\ 0
         \end{array}\right) = \\
   & = (\cos(nt),\; \sin(nt),\; 0) \R_z(-M_{02})\R_x(-i_2)\R_z(\Delta\Omega)
       \R_x(i_1)\R_z(M_{01})\left(\begin{array}{c}
         \cos(nt)\\ \sin(nt) \\ 0
         \end{array}\right)
  \end{aligned}
\]
where $\Delta\Omega=\Omega_1-\Omega_2$. Finding the minimum $\gamma(t)$ over
an orbital period is equivalent to maximizing
$\cos(\gamma(t))$ for $t\in[0,T]$, or more simply, to maximizing
\[
  \begin{aligned}
  &\left(\cos( \beta),\; \sin(\beta),\; 0\right)
  \R_z(-M_{02})\R_x(-i_2)\R_z(\Delta\Omega)
  \R_x(i_1)\R_z(M_{01})\left(\begin{array}{c}
    \cos(\beta)\\ \sin(\beta) \\ 0 \end{array}\right) = \\
  & \quad\, = \left(\cos(\beta+M_{02}),\; \sin(\beta+M_{02}),\; 0\right)
  \R_x(-i_2)\R_z(\Delta\Omega)
  \R_x(i_1)\left(\begin{array}{c}
    \cos(\beta+M_{01})\\ \sin(\beta+M_{01}) \\ 0 \end{array}\right)
  \end{aligned}
\]
for $\beta=nt\in[0,2\pi]$. Changing variables $\beta'=\beta+M_{02}$, this expression becomes
\begin{align}
  & \left(\cos(\beta'),\; \sin(\beta'),\; 0\right)
  \R_x(-i_2)\R_z(\Delta\Omega)
  \R_x(i_1)\left(\begin{array}{c}
    \cos(\beta'+M_{01}-M_{02})\\ \sin(\beta'+M_{01}-M_{02}) \\ 0
    \end{array}\right) = \nonumber \\
  & \qquad\, = \left(\cos(\beta'),\; \sin(\beta'),\; 0\right)
  \R_x(-i_2)\R_z(\Delta\Omega)
  \R_x(i_1)\R_z(\Delta M_0)\left(\begin{array}{c}
    \cos(\beta')\\ \sin(\beta') \\ 0
    \end{array}\right) \label{eq1}
\end{align}
where $\Delta M_0=M_{01}-M_{02}$. Assume that the product of the four
rotation matrix in the expression above is
\[
\R_x(-i_2)\R_z(\Delta\Omega)\R_x(i_1)\R_z(\Delta M_0) =
\left(\begin{array}{ccc}a & b & * \\ c & d & * \\ * & * & *\end{array}\right)
\]
for some $a,b,c,d\in\mathbb{R}$. The entries marked with an asterisk
are not relevant, since they will later be multiplied by
zeros. The expression~\eqref{eq1} that we want to maximize 
can be rewritten as
\begin{align}
 & \left(\cos(\beta'),\; \sin(\beta'),\; 0\right)
   \left(\begin{array}{ccc}a & b & * \\ c & d & * \\ * & * & *\end{array} 
   \right)\left(\begin{array}{c}\cos(\beta')\\ \sin(\beta') \\ 0
    \end{array}\right) = \nonumber \\
 & \qquad\, =  a\cos^2(\beta') + (b+c)\cos(\beta')\sin(\beta') + d\sin^2(\beta') = \nonumber \\
 & \qquad\, = \frac{a+d}2 + \frac{a-d}2\cos(2\beta') + \frac{b+c}2\sin(2\beta') \label{eq2}
\end{align}
where $\beta'$ ranges from $0$ to $2\pi$.

In general, the maximum of an expression of the form $x\cos(\theta)+y\sin(\theta)$ for $\theta\in[0,2\pi]$ happens when the unit vector
$(\cos(\theta),\sin(\theta))$ is aligned (same direction) with $(x,y)$,
i.e. when \[\left(\cos(\theta),\;\sin(\theta)\right)=\left(\frac{x}{\sqrt{x^2+y^2}},\;\frac{y}{\sqrt{x^2+y^2}}\right).\] At this particular point, the value of
the function is \[x\cdot \frac{x}{\sqrt{x^2+y^2}} + y \cdot \frac{y}{\sqrt{x^2+y^2}}= \sqrt{x^2+y^2}.\]

Applying this idea to our maximization problem, we get
\[
  \max_{t\in [0,T]}\cos(\gamma(t)) = \frac{a+d}2 + \frac{\sqrt{(a-d)^2+(b+c)^2}}2
\]
which translates into
\begin{equation}\label{min-ang-formula}
  \min_{t\in[0,T]}\gamma(t) = \arccos\left(\frac{a+d}2 + \frac{\sqrt{(a-d)^2+(b+c)^2}}2\right)
\end{equation}
for the minimum angular distance between the two satellites. The expressions
for $a,b,c,d$ in terms of $i_1,i_2,\Delta\Omega,\Delta M_0$ can be
easily obtained by multiplying the four rotation matrices. The following
pseudocode shows all these formulas.

\bigskip

\begin{algorithm}[!ht]
\caption{Minimal angular distance between satellites in circular orbits of equal radius}\label{ang-dist}
\begin{algorithmic}[1]
\Input the orbital parameters $(i_1, \Omega_1, M_{01})$ and
$(i_2, \Omega_2, M_{02})$ of the two satellites.
\Output the minimum angular distance $\alpha_{min}$ between the satellites
in an orbital period.
\Procedure{MinAngDist}{$i_1,\Omega_1,M_{01},i_2,\Omega_2,M_{02}$}
\State $\Delta\Omega = \Omega_1 - \Omega_2$
\State $\Delta M_0   = M_{01}   - M_{02}$
\State $a =  \cos(\Delta\Omega)\cos(\Delta M_0) -
             \sin(\Delta\Omega)\cos(i_1)\sin(\Delta M_0)$
\State $b = -\cos(\Delta\Omega)\sin(\Delta M_0) -
             \sin(\Delta\Omega)\cos(i_1)\cos(\Delta M_0)$
\State $c =  \cos(i_2)\sin(\Delta\Omega)\cos(\Delta M_0) +
             \cos(i_2)\cos(\Delta\Omega)\cos(i_1)\sin(\Delta M_0)+$

       $  +  \sin(i_2)\sin(i_1)\sin(\Delta M_0)$
\State $d = -\cos(i_2)\sin(\Delta\Omega)\sin(\Delta M_0) +
             \cos(i_2)\cos(\Delta\Omega)\cos(i_1)\cos(\Delta M_0) +$

       $  +  \sin(i_2)\sin(i_1)\cos(\Delta M_0)$
\State $e=0.5\left( a+d + \sqrt{(a-d)^2+(b+c)^2}\right)$
  \Comment{$e$ is guaranteed to be in $[-1,1]$}
\State $\alpha_{min}=\arccos(e)$
  \Comment{$\arccos()$ returns a value in $[0,\pi]$}
\State \textbf{return} $\alpha_{min}$
\EndProcedure
\end{algorithmic}
\end{algorithm}

The total count of floating point operations used in
algorithm~\ref{ang-dist} is: $8$ standard trigonometric functions
($\sin$ and $\cos$), $1$ inverse
trigonometric ($\arccos$), $1$ square root, $23$ multiplications, and $13$
additions and subtractions.

\bigskip

The operation count above is a bit naive, since it is clear that many
computations are repeated in several places. For instance, the product
$\sin(i_1)\sin(i_2)$ is computed in lines $6$ and $7$. A clever reordering
of the operations, as shown in Algorithm~\ref{fast-ang-dist}, can reduce
the number of multiplications to only $17$.

\bigskip

The method can also be easily parallelized. On a powerful enough machine,
the lines 2--3, 4--11, 12--17, 18--21 of Algorithm~\ref{fast-ang-dist} can
be processed in parallel (see the annotations in the pseudocode), since
there are no dependencies in either group. We have left all this kind of
parallelization (which depends strongly on the type of processor used) to
the compiler.

\bigskip

An implementation in C of Algorithm~\ref{fast-ang-dist} on a modern
computer (Asus UX430, Ubuntu 18.04.4 LTS) compiled with gcc $7.5.0$ runs,
on a single core, at a rate of $4.79\cdot 10^6$ calls per second (double
precision) and $1.24\cdot 10^7$ calls per second (single precision). The
processor of this machine is an Intel Core i5-7200U.

\bigskip

A OpenCL implementation, without any special optimization, running on
the GPU of the same machine (Intel Gen9 HD Graphics NEO), can process
approximately $4.90\cdot 10^7$ calls per second (double precision) and
$6.58\cdot 10^7$ calls per second (single precision). The timing not
only includes the computation time, but also the time needed to move
the data from the main memory to the GPU memory and vice versa. In
double precision, the GPU is more than 10 times faster than a CPU core.
A hand optimized version runs approximately four times faster.

\bigskip

\begin{algorithm}[!ht]
\caption{Minimal angular distance between satellites in circular orbits
of equal radius -- Optimized -- Annotated for parallelization}
\label{fast-ang-dist}
\begin{algorithmic}[1]
\Input the orbital paramet\tikzmark{right1}ers $(i_1, \Omega_1, M_{01})$ and
$(i_2, \Omega_2, M_{02})$ of the two satellites.
\Output the minimum angular distance $\alpha_{min}$ between
\tikzmark{right2} the satellites in an orbital period.
\Procedure{MinAngDist}{$i_1,\Omega_1,M_{01},i_2,\Omega_2,M_{02}$}
\State $\Delta\Omega = \Omega_1 - \Omega_2$ \tikzmark{top1}
\State $\Delta M_0 = M_{01} - M_{02}$       \tikzmark{bot1}
\State $C\Omega=\cos(\Delta\Omega)$         \tikzmark{top2}
\State $S\Omega=\sin(\Delta\Omega)$
\State $CM_0=\cos(\Delta M_0)$
\State $SM_0=\sin(\Delta M_0)$
\State $CI_1=\cos(i_1)$
\State $SI_1=\sin(i_1)$
\State $CI_2=\cos(i_2)$
\State $SI_2=\sin(i_2)$                     \tikzmark{bot2}
\State $aux_1=C\Omega \cdot CM_0$           \tikzmark{top3}
\State $aux_2=S\Omega \cdot CM_0$
\State $aux_3=C\Omega \cdot SM_0$
\State $aux_4=S\Omega \cdot SM_0$
\State $aux_5=CI_1 \cdot CI_2$
\State $aux_6=SI_1 \cdot SI_2$              \tikzmark{bot3}
\State $a =  aux_1 - aux_4\cdot \cos(i_1)$  \tikzmark{top4}
\State $b = -aux_3 - aux_2\cdot \cos(i_1)$
\State $c =  aux_2 \cdot \cos(i_2) + aux_3\cdot aux_5 + SM_0\cdot aux_6$
\State $d = -aux_4 \cdot \cos(i_2) + aux_1\cdot aux_5 + CM_0\cdot aux_6$
                                            \tikzmark{bot4}
\State $e=0.5\left( a+d + \sqrt{(a-d)^2+(b+c)^2}\right)$
  \Comment{$e$ is guaranteed to be in $[-1,1]$}
\State $\alpha_{min}=\arccos(e)$
  \Comment{$\arccos()$ returns a value in $[0,\pi]$}
\State \textbf{return} $\alpha_{min}$
\EndProcedure
\end{algorithmic}
\AddNote{top1}{bot1}{right1}{parallel group 1}
\AddNote{top2}{bot2}{right1}{parallel group 2}
\AddNote{top3}{bot3}{right1}{parallel group 3}
\AddNote{top4}{bot4}{right2}{parallel group 4}
\end{algorithm}

To our knowledge, the best method known up to now for computing the
minimum angular distance $\alpha_{min}$ between two satellites in
circular orbits of the same radius without propagation is a formula
proven by Speckman, Lang, and Boyce in~\cite{speckman}.
\begin{equation}\label{speckman-formula}
  \begin{aligned}
  \alpha_{min} &= 2\left|\arcsin\left(\sqrt{\frac{1+\cos(i_1)\cos(i_2)+\sin(i_1)\sin(i_2)\cos(\Delta\Omega)}{2}}\sin\left(\frac{\Delta F}{2}\right)\right) \right| \\
  \Delta F &= \Delta M_0 -2\arctan\left(-\tan\left(\frac{\Delta\Omega}{2}\right)\frac{\cos\left(\frac{i_1+i_2}{2}\right)}{\cos\left(\frac{i_1-i_2}{2}\right)}\right)
  \end{aligned}
\end{equation}
This formula uses $9$ trigonometric functions ($\sin$, $\cos$, $\tan$),
$2$ inverse trigonometric ($\arcsin$, $\arctan$), $1$ square root,
$9$ multiplications and divisions, and $5$ additions and subtractions.
In comparison, our method uses fewer trigonometric functions and inverse
trigonometric functions, but more arithmetic operations. On a machine
where the trigonometric functions dominate the computation, it is
reasonable to expect that our method would be faster. Indeed, an
implementation in C of the formula above can process approximately
$3.69\cdot 10^6$ calls per second (double precision) and $7.11\cdot 10^6$
calls per second (single precision), under the same conditions we tested
our formula. Compared to that, our method can process $30\%$ more calls
per second in double precision and $74\%$ more in single precision.
Finally, to validate the accuracy of our method, we tested
formulas~\eqref{min-ang-formula} and~\eqref{speckman-formula} on a random
sample of $10^7$ cases and verified that they return values within
$2.15\cdot 10^{-10}$ of each other in double precision and
$5.79\cdot 10^{-4}$ in single precision.

\section{Fast evaluation of Flower Constellations}\label{sec3}

A 2D Lattice Flower Constellation (see~\cite{2dlfc}) is defined by three
integer parameters $N_o\geq 1$, $N_{so}\geq 1$, and $0\leq N_c<N_o$, and
the orbital parameters of a reference satellite
$(a,e,incl,\omega,\Omega,M_0)$.
The constellation has $N_{sat}=N_oN_{so}$ satellites denoted $Sat_{ij}$
whose orbital elements are $(a,e,incl,\omega,\Omega_{ij}, M_{0,ij})$, where
\[
  \Omega_{ij} = \Omega + 2\pi \frac{i}{N_o}
  \qquad
  M_{0,ij}     = M_0    + 2\pi \frac{jN_o-iN_c}{N_{sat}}
\]
for $i=0,\ldots,N_o-1$ and $j=0,\ldots,N_{so}-1$. The indices $i$ and $j$
will always be regarded as integers modulo $N_o$ and $N_{so}$, respectively.
For instance $Sat_{N_o+3,4}=Sat_{3,4}$. The first four orbital
elements $(a,e,incl,\omega)$ are common to all satellites. The reference satellite is
$Sat_{00}$. In this section, we show how to evaluate efficiently the
minimum angular distance between any pair of satellites of a FC.

\bigskip

The number of possible FCs with a maximum given number of satellites $k$ is
\[
  \sum_{N_o=1}^{k} N_o\left\lfloor\frac{k}{N_o}\right\rfloor
\]
since for any possible $N_o$, the possible values of $N_{so}$ are the
positive integers such that $N_oN_{so}\leq k$, and the possible values
for $N_c$ are the integers from $0$ to $N_o-1$. Each term of the sum above
is bounded above by $k$ and below by $k/2$. Therefore, the number of FCs
with $N_{sat}\leq k$ is between $k^2/2$ and $k^2$.

\bigskip

A simple but inefficient method to evaluate a FC is to compute the
value of $\alpha_{min}$ for each pair of satellites of the constellation
and return the minimum of those values. This method requires
$N_{sat}(N_{sat}-1)/2$ calls to the formula to compute $\alpha_{min}$.
While this method might work well for a single FC, it becomes too costly
when one needs to evaluate all FCs with $N_{sat}\leq k$. Indeed, the number
of calls to the formula for $\alpha_{min}$ would be
\[
  \sum_{N_o=1}^{k}\sum_{N_{so}=1}^{\lfloor k/N_o\rfloor}
    \frac{N_o (N_oN_{so}) (N_oN_{so}-1)}{2}\approx \frac{k^4}{6}.
\]
Even for FCs with circular orbits, where we can use
Algorithm~\ref{fast-ang-dist}, it will take more than three months to
process the case $k=10^4$ using the same computer and the optimized OpenCL
code mentioned in Section~\ref{sec2}. Nevertheless, as we show below, it
is possible to use properties of the FCs to reduce this time significantly.

\bigskip

The main properties of FCs are their symmetries.
If constellations are
regarded as ``3d-movies'' showing the motion of the satellites, FCs
are invariant under the following two operations:
\begin{itemize}
\item $R_t(T/N_{so})$: shifting the time scale of the movie by $T/N_{so}$,
where $T$ is the period of the satellites. If the original movie and the
movie with the time scale shifted are projected together, a viewer will
see exactly the same frame. Of course, each satellite will occupy the position
of another, but the overall configuration will be the same. Under this
operation, the satellite $Sat_{i,j}$ of the original FC will occupy the
location of the satellite $Sat_{i,j+1}$ of the other.
\item $R_z(2\pi/N_o)R_t(N_cT/N_{sat})$: this operation combines a rotation
about the $z$-axis and a shift of the time scale. Under this operation,
the satellite $Sat_{i,j}$ of the original FCs corresponds to $Sat_{i+1,j}$
of the rotated FCs.
\end{itemize}
These symmetries can be combined to create more complicated ones. For
instance, for any $\delta_i$ and~$\delta_j$, there is an operation that
maps $Sat_{i,j}$ of the initial FC into $Sat_{i+\delta_i,j+\delta_j}$
of the transformed~FC.

\bigskip

A nice consequence of the symmetries of a FC is that it is possible to
evaluate a FC by only considering the angular distances between the
reference satellite and the other $N_{sat}-1$. Indeed, the minimum
angular distance between $Sat_{i_1,j_1}$ and $Sat_{i_2,j_2}$ is exactly
the same as between $Sat_{0,0}$ and $Sat_{i_2-i_1,j_2-j_1}$. Applying
this trick, we can evaluate all FCs with $N_{sat}\leq k$ with
\[
  \sum_{N_o=1}^{k}\sum_{N_{so}=1}{\lfloor k/N_o\rfloor} N_o (N_oN_{so}-1)
  \approx \frac{k^3}{2}
\]
calls to the routine that computes $\alpha_{min}$ between a pair
of satellites. For FCs with circular orbits, the case $k=10^4$ requires
$40$ minutes of computation.

\bigskip

The notion of distance is clearly symmetrical, i.e. the distance between
$Sat_{i_1,j_1}$ and $Sat_{i_2,j_2}$ is the same as between $Sat_{i_2,j_2}$
and $Sat_{i_1,j_1}$. However, according to the result of the previous
paragraph, these two distances correspond to the ones from the reference
satellite to $Sat_{i_1-i_2,j_1-j_2}$ and $Sat_{i_2-i_1,j_2-j_1}$,
respectively. Therefore, these two distances must be equal, so only
one has to be computed. Due to the modular nature of the indices $i$ and
$j$, only the distances from the reference satellite to satellites
$Sat_{ij}$ with $i\leq\lfloor N_o/2\rfloor$ have to be computed. A
pseudocode showing how to implement this idea is given in
Algorithm~\ref{fc-ang-dist}. This trick reduces the computation
time in half, i.e. to $k^3/4$ calls, so the case $k=10^4$ would only take $20$ minutes.

\bigskip

\begin{algorithm}[!ht]
\caption{Minimal angular distance for a FC with circular orbits}
\label{fc-ang-dist}
\begin{algorithmic}[1]
\Input the parameters $N_o$, $N_{so}$, $N_c$ that define the FC.
\Input the orbit inclination $incl$.
\Output the minimum angular distance $\alpha_{min}$ between any pair of
satellites of the FC in an orbital period.
\Procedure{FCMinAngDist}{$N_o, N_{so}, N_c, incl$}
\State $N_{sat}=N_o\cdot N_{so}$
\State $\alpha_{min}=2\pi$
\For{$i=0,\ldots,\lfloor N_o/2\rfloor$}
\For{$j=0,\ldots,N_{so}-1$}
\If{$i=0$ and $j=0$}\label{if-cond}
  \Comment{do not compare the reference satellite to itself}
\State \textbf{go to line} \ref{xxx}
\EndIf
\State $\Omega = 2\pi\cdot i/N_o$
\State $M_0    = 2\pi\cdot (j\cdot N_o-i\cdot N_c)/N_{sat}$
\State $\alpha = \textsc{MinAngDist}(incl,\, 0.0,\, 0.0,\, incl,\, \Omega,\, M_0)$
\If{$\alpha < \alpha_{min}$}
\State $\alpha_{min}=\alpha$
\EndIf
\EndFor \label{xxx}
\EndFor
\State \textbf{return} $\alpha_{min}$
\EndProcedure
\end{algorithmic}
\end{algorithm}

The condition in line~\ref{if-cond} of Algorithm~\ref{fc-ang-dist} can
be improved a little bit.
\begin{equation}\label{line}
  \textbf{if}\;(i=0\;\text{and}\;j\leq\lfloor N_{so}/2\rfloor)\;\text{or}\;
  (i=N_o/2\;\text{and}\;j > \lfloor N_{so}/2\rfloor)\;\textbf{then}
\end{equation}
Instead of only removing the reference satellite, it is possible to
remove all the satellites such that $i=0$ and
$j\leq\lfloor N_{so}/2\rfloor$. The distance between any such satellite
$Sat_{0,j}$ and the reference satellite is equal to the distance between
$Sat_{0,-j}=Sat_{0,N_{so}-j}$ and the reference satellite. It is clearly
impossible that both $Sat_{0,j}$ and $Sat_{0,N_{so}-j}$ are excluded from
the search by the new condition.
Similarly, in the case where $N_o$ is an even integer, the satellites
with $i=N_o/2$ and $j>\lfloor N_{so}/2\rfloor$ can be excluded without
losing any information. Doing this replacement will bring down the number
of calls to Algorithm~\ref{fast-ang-dist} to exactly
$\lfloor N_{sat}/2\rfloor$.

\bigskip

So far, we have only dealt with the integer parameters of the FCs. In the
case of FCs with circular orbits, the reference satellite has orbital
parameters $(a,e=0,incl,\omega=0,\Omega,M_0)$. The value of the semimajor
axis $a$, which in this case is the radius of the orbit, does not affect
the angular distance. The value of $\Omega$ and $M_0$ do not affect the
distance either, since a constellation with non-zero values of $\Omega$
and $M_0$ can be transformed into one with $\Omega=M_0=0$ by applying
the transformation $R_z(\Omega)R_t(T\cdot M_0/(2\pi))$. The only parameter
that matters is the value of the inclination. Since the value is a real
number, no exhaustive search is possible. A discretization of this value
in $l$ possibilities will bring the running time to $C/4\cdot k^3\cdot l$,
where $C$ is the average time per call of the routine implementing
Algorithm~\ref{fast-ang-dist}.

\bigskip

Assume now that $N_o$ and $N_c+N_{so}$ are both divisible by two. In this
case, Algorithm~\ref{fc-ang-dist} always returns $\alpha_{min}=0$, i.e. the
FC has collisions. The reason is that in the main loop, when $i=N_o/2$ and
$j=(N_c+N_{so})/2$, the values of $\Omega$ and $M_0$ become
\[
  \Omega = 2\pi\frac{N_o/2}{N_o}=\pi \qquad
  M_0 = 2\pi\frac{\frac{N_c+N_{so}}{2}N_o-\frac{N_o}{2}N_c}{N_{sat}}=\pi
\]
and $\textsc{MinAngDist}(incl,0,0,incl,\pi,\pi)=0$. If these FCs are
discarded, which represent about $25\%$ of the total number of FCs with
$N_{sat}\leq k$, then the total cost of computation reduces to
$3C/16\cdot k^3\cdot l$. For a single inclination, the case $k=10^4$
would take only $15$ minutes.

\section{Conclusions}

Algorithm~\ref{fast-ang-dist} computes the minimum angular distance,
measured from the center of the Earth, between two given satellites
moving in circular orbits of the same radius, during an orbital period.
The method is easy to implement and does not perform any propagation
of the satellites. In terms of floating-point operations, it uses $8$
trigonometric functions ($\sin$ and $\cos$), $1$ inverse trigonometric
function ($\arccos$), $1$ square root, $17$ multiplications, and $13$
additions and subtractions. On a modern laptop, a single CPU core is
able to process up to $4.79\cdot10^6$ calls per second in double
precision. The method is highly parallelizable, with a dependency chain
of length~$6$.

\bigskip

Algorithm~\ref{fc-ang-dist} computes the minimum angular distance
between all satellites of a FC with circular orbits. If the
improved~\eqref{line} is used instead of line~\ref{if-cond}, only
$\lfloor N_{sat}/2\rfloor$ pairs of satellites are evaluated with
Algorithm~\ref{fast-ang-dist}. This improves the naive method of
testing every pair of satellites by a factor of $2(N_{sat}-1)$.

\bigskip

The total number of FCs with $N_{sat}\leq k$ is between $k^2/2$
and $k^2$. Running Algorithm~\ref{fc-ang-dist} for all such FC
and $l$~different inclinations requires $k^3/4\cdot l$ calls to
Algorithm~\ref{fast-ang-dist}.

\bigskip

FCs with $N_o$ and $N_{so}+N_c$ both divisible by two, always have
collisions, i.e. their minimum angular distance is zero. Pruning these
cases speeds up the computation of the previous paragraph to only
$3k^3/16\cdot l$ calls to Algorithm~\ref{fast-ang-dist}.
On a modern Nvidia Tesla K40c GPU, a hand optimized implementation of
this method can process the case $k=10^4$ and $l=180$ in only $3.35$
hours.

\section*{Acknowledgments}

Mart\'\i n Avenda\~no is partially supported by grant MTM2016-76868-C2-2-P
(Ministry of Science and Innovation, Spain).
David Arnas is partially supported by grant ESP2017-87113-R (Ministry of
Science and Innovation, Spain).
All authors are also funded by MISTI Global Seed Funds (La Caixa
Foundation).

\section*{References}

\bibliography{efficient-search}

\begin{thebibliography}{1}
\expandafter\ifx\csname url\endcsname\relax
  \def\url#1{\texttt{#1}}\fi
\expandafter\ifx\csname urlprefix\endcsname\relax\def\urlprefix{URL }\fi
\expandafter\ifx\csname href\endcsname\relax
  \def\href#1#2{#2} \def\path#1{#1}\fi

\bibitem{stmfc}
D.~Arnas, M.~Lifson, R.~Linares, M.~Avenda{\~n}o, {Low Earth Orbit slotting for
  Space Traffic Management using Flower Constellation Theory}, in: AIAA Scitech
  2020 Forum, 2020, p. 0721.
\newblock \href {http://dx.doi.org/10.2514/6.2020-0721}
  {\path{doi:10.2514/6.2020-0721}}.

\bibitem{leemorav16}
S.~Lee, M.~Avenda{\~n}o, D.~Mortari, Uniform and weighted coverage for large
  lattice flower constellations, in: Advances in the Astronautical Sciences
  Astrodynamics, Vol. 156, American Astronautical Society, Univelt,
  Incorporated, P.O. Box 28130, San Diego, California 92198, 2015, pp.
  3633--3648, iSBN: 978-0-87703-629-6.

\bibitem{2dlfc}
M.~Avenda{\~n}o, J.~Davis, D.~Mortari, The 2d lattice theory of flower
  constellations, Celestial Mechanics and Dynamical Astronomy 116 (2013)
  325--–337.
\newblock \href {http://dx.doi.org/10.1007/s10569-013-9493-8}
  {\path{doi:10.1007/s10569-013-9493-8}}.

\bibitem{arnas18}
D.~Arnas, D.~Casanova, E.~Tresaco, {2D Necklace Flower Constellations}, Acta
  Astronautica 142 (2017) 18--28.
\newblock \href {http://dx.doi.org/10.1016/j.actaastro.2017.10.017}
  {\path{doi:10.1016/j.actaastro.2017.10.017}}.

\bibitem{speckman}
L.~Speckman, T.~Lang, W.~Boyce, An analysis of the line of sight vector between
  two satellites in common altitude circular orbits, in: Astrodynamics
  Conference, Vol.~24, American Institute of Aeronautics and Astronautics Inc,
  AIAA, Portland, OR., 1990, pp. 866--874.
\newblock \href {http://dx.doi.org/10.2514/6.1990-2988}
  {\path{doi:10.2514/6.1990-2988}}.

\end{thebibliography}

\end{document}